\documentstyle[12pt,epsf]{article}
\def\Po{P^{{\rm I}}}
\def\Pt{P^{{\rm II}}}
\begin{document}
\title{{\bf Persistence in the Voter model: continuum reaction-diffusion 
 approach}}
\author{M Howard\dag, C Godr\`eche\ddag\S}
\date{
{\small\noindent\dag Center for Chaos and Turbulence Studies, The Niels Bohr 
Institute, Blegdamsvej 17, 2100 Copenhagen \O, Denmark \\}
{\small\noindent\ddag Service de Physique de l'\'Etat Condens\'e, CEA-Saclay, 
91191 Gif sur Yvette, France \\}
{\small\noindent\S Laboratoire de Physique Th\'eorique et Mod\'elisation, 
Universit\'e de Cergy-Pontoise, France \\}}
\maketitle

\begin{abstract}
We investigate the persistence probability in the Voter model for 
dimensions $d\geq 2$. This is achieved by mapping the Voter model 
onto a continuum reaction--diffusion system. 
Using path integral methods, we  compute the persistence probability
$r(q,t)$, where $q$ is the number of ``opinions'' in the original Voter 
model. We find $r(q,t)\sim\exp[-f_2(q)(\ln t)^2]$ in $d=2$; $r(q,t)\sim
\exp[-f_d(q)t^{(d-2)/2}]$ for $2<d<4$; $r(q,t)\sim\exp[-f_4(q)t/\ln t]$
in $d=4$; and $r(q,t)\sim\exp[-f_d(q)t]$ for $d>4$. The results of our 
analysis are checked by Monte Carlo simulations.
\end{abstract}


\noindent
The Voter model is a simple stochastic model which exhibits interesting
dimension dependent properties \cite{Liggett}.
Whilst, in one dimension, it is equivalent to the Glauber-Ising model at zero
temperature, its properties depart from this model in higher dimensions.
On each site of a $d-$dimensional lattice, opinions of a voter, or values of a
spin
$\sigma=1, 2,\ldots,q$, are initially distributed randomly.
Between $t$ and $t+dt$ a site is picked at random.
The voter on this site takes the opinion of one of its $2d$ neighbours, 
also chosen at random.
This model is equivalent to a system of coalescing random
walkers \cite{Liggett}. 
Therefore, due to the recurrence properties of random walks, the
interactions between walkers are strong in low dimensions ($d\le2$), and
progressively decrease in higher dimensions. 
As a consequence, for $d\le2$ a consensus of opinion is approached as
$t\to\infty$, or in other words, the system coarsens. This
contrasts with the behaviour in higher
dimensions where disagreements persist indefinitely \cite{Liggett}.

In this letter we compute the persistence probability $r(q,t)$, i.e.
the probability that a given voter
never changed his opinion up to time $t$, in
arbitrary dimension $d\geq 2$, using field theoretical methods. 
Persistence probabilities have been investigated for a number of coarsening
systems, including the Glauber-Ising chain \cite{derrida1, derrida2,derrida3},
exhibiting in all cases an algebraic decay with time.
By contrast, it was found numerically that the persistence probability
of the $2d$ Voter model with $q=2$ had the behaviour
$\exp[-{\rm const.}(\ln t)^2]$ \cite {bennaim}. 
This change in behaviour with
dimension motivated the present work. 
Field theoretical methods are
well adapted for the study of the role of dimension in simple physical systems, hence
their use is natural in the present context.

The first step for the study of persistence consists in mapping the Voter model
onto its dual particle system, i.e. a
system of particles on a  $d-$dimensional lattice, with the reaction 
$A+A\to A$,
fermionic occupation rules and particle injection at the origin. 
As soon as one particle leaves that site another
particle is immediately injected there in its place. 
This reaction-diffusion model is hereafter referred to as model I.
Defining $Q(m,t)$ as the probability that, after a time $t$ there are $m$
particles in the system, one has 
$r(q,t)=\sum_{m=1}^{\infty}Q(m,t)~q^{1-m}$ \cite{derrida2,derrida3}.
After a time $t$ the injected particles will have penetrated a distance of
order $t^{1/2}$ from the source. 
At greater distances there are almost no particles present, whereas behind the
front a steady state is reached. 
Hence in order to calculate the probability $r(q,t)$, one can proceed as
follows: (i) Compute
\begin{equation}
s(q,L)=\sum_{m=1}^N\Po(m,L)~q^{1-m},\label{sql}
\end{equation}
where $\Po(m,L)$ is the steady state probability of finding $m$ 
particles within
a distance $L$ of the point of injection (for $t^{1/2}\gg L$), and 
where $N={\cal O}(L^d)$ is the number of sites within that radius. 
(ii) Now compute the persistence probability using finite size scaling:
$r(q,t)\sim s(q,L\sim t^{1/2})$.

Unfortunately a computation of the $\Po(m,L)$ is still too difficult. 
Therefore we replace model I with a slightly modified variant, which we 
will call model II. We now have multiple (bosonic) site occupancy 
with a finite reaction rate $\lambda$, and particle input at the origin 
with a  rate $J$. However the relevant properties of both models
will turn out to be dominated by their (low density) behaviour 
far from the injection
point. This means that, in the large $L$ limit, the distributions $\Po$ 
and $\Pt$ actually become very similar, hence   
model II can be used for the persistence calculation.   
Evidence for this claim will be given later.

The mean field equation for model II, with particle density $\rho(x,t)$, is
given by (for details see \cite{cheng,krapivsky})
\begin{equation}
\partial_t \rho=D\nabla^2 \rho-\lambda \rho^2+J\delta^{d}(x), \label{mfe}
\end{equation}
where $D$ is the diffusion constant. 
Well behind the diffusive front, but far from the injection point,
we can insert a steady-state power law ansatz into (\ref{mfe}), giving
\begin{equation}
\label{mfs}
\rho(x)\sim\cases{2(4-d)D/(\lambda x^2), & ($d<4$) \cr
2D/(\lambda x^2\ln x), & ($d=4$) \cr
{\rm const.}\, ~x^{2-d}, & ($d>4$).} 
\end{equation}
Note that for $d\leq 4$, the asymptotic form of $\rho(x)$ does not depend 
on $J$, an indication of the similarity between models I and II.
The change in behaviour at $d=4$ is due to the role of the reaction term 
in (\ref{mfe}). 
For large $x$, this reaction term is marginal in $d=4$, whereas for $d>4$,
it can be asymptotically ignored, the density instead decaying in
accordance with Laplace's equation. 
Hence $d=4$ is, in a sense, a critical dimension. 
Nevertheless the behaviour of the average density around $d=4$
can be described adequately within mean field theory. However
in low spatial dimensions
($d\leq d_c=2$), the mean field equation (\ref{mfe}) is no longer correct,
since it does not include fluctuation effects. 
We therefore map model II to a field theory which systematically
includes these fluctuations, and permits an analysis of the probability
distribution $\Pt(m,L)$ in arbitrary dimension.
Following the work of \cite{doi,peliti},
and using the formalism of \cite{lee}, the master equation 
for model II may be rewritten as a second quantized Hamiltonian
\begin{equation}
\hspace{-.4in}
\hat H=-D\sum_{n.n.}\{\hat a_i^{\dagger}(\hat a_j
-\hat a_i)\}-\lambda
\sum_i\{(1-\hat a_i^{\dagger})\hat a_i^{\dagger}\hat a_i^2\}-J
\{\hat
a_{1}^{\dagger}-1\}.
\end{equation}
The first term describes diffusion of the $A$ particles, the second 
term describes their coagulation reaction, and the final term results 
from stochastic (Poissonian) particle injection at the origin (site $1$). 
Defining $P(n_1,\ldots,n_N)$ to be the probability that the system is in a
configuration $\{n_1,\ldots,n_N\}$ ($n_i$ is the occupation number
for site $i$), then the steady state probability 
$\Pt(m,L)$ of finding $m$ particles in the system is
\begin{equation}
\Pt(m,L)=\sum_{\{n\}} P(n_1,\ldots,n_N)~\delta(\sum_j n_j, m).
\end{equation}
In the second quantized formalism, where $|\Psi\rangle$
is the system state ket, this becomes
\begin{eqnarray}
& & 
\Pt(m,L)=\langle 0|\exp(\sum_i \hat a_i)~\delta(\sum_j n_j,m)~|\Psi
\rangle \\
& & \hspace{4.3em} = {1\over m!}\sum_{n_1\ldots n_N \atop \sum_j n_j=m}~{
m!\over n_1!\ldots n_N!}~\langle 0| \hat a_1^{n_1}\dots \hat
a_N^{n_N}|\Psi\rangle.\nonumber
\end{eqnarray}
Identifying this last expression as a multinomial expansion
yields
\begin{equation}
\Pt(m,L)=\langle 0|{1\over m!}(\sum_i \hat a_i)^m|\Psi\rangle.
\end{equation}
Hence the generating function for the probabilities $\Pt(m,L)$ is given
by
\begin{equation}
\hspace{-.4in}
G(y)=\sum_{m=0}^{\infty} y^m \Pt(m,L)=
\langle 0|\exp(y\sum_i \hat a_i)|\Psi\rangle
=\langle |\exp[(y-1)\sum_i \hat a_i]|\Psi\rangle,
\label{genlattice}
\end{equation}
with the definition $\langle|\equiv\langle 0|\exp(\sum_i\hat a_i)$.
Performing the mapping to a path integral \cite{doi,peliti,lee} yields 
the continuum action
\begin{equation}
S=\int d^dx\left.\left.\int dt~\right(\bar a(\partial_t-\nabla^2)a+
\lambda\bar aa^2+\lambda\bar a^2a^2-J\bar a\delta^d(x)\right),
\label{action}
\end{equation}
where we have absorbed the diffusion constant $D$ into a rescaling of 
time and of the couplings $\lambda$ and $J$. Furthermore the continuum 
limit of (\ref{genlattice}) is
\begin{equation}
G(y)=\left\langle\exp\left[(y-1)\int d^dx~a(x)\right]\right\rangle,
\label{genel}
\end{equation}
with the average performed with respect to the action $S$. 
{}From (\ref{genlattice}) we have
\begin{equation}
\Pt(m,L)=\left.{1\over m!}{d^m\over dy^m} G(y)\right|_{y=0}
\end{equation}
which leads to the continuum expression
\begin{equation}
\hspace{-.5in}\Pt(m,L)={1\over m!}\left\langle\int d^dx_1~a(x_1) 
\ldots \left.\left.\int d^dx_m ~a(x_m) \exp\right[-\int d^dx~a(x)\right]
\right\rangle,\label{conpii}
\end{equation}
and hence, from (\ref{sql}),
\begin{equation}
s(q,L)\sim G(q^{-1})=\left\langle\exp\left[-(1-q^{-1})\int d^dx~a(x)
\right]\right\rangle, \label{qcalc}
\end{equation}
where we have used the equivalence between models 
I and II in the large $L$ limit, mentioned previously.

The reaction rate $\lambda$ 
has naive scaling dimension $k^{2-d}$. 
This predicts a critical
dimension $d_c=2$ at or below which we must renormalize the theory.
A very similar case has been studied in \cite{howard1}. Hence
the dimensionless renormalized coupling evaluated at the normalization point
$s=\kappa^2$ is given by $g_R=\kappa^{-\epsilon}\lambda_R$, where 
$\epsilon=2-d$ and
\begin{equation}
\lambda_R=\lambda(1+\lambda I_d(\kappa))^{-1},\qquad
I_d(\kappa)=\left. 2^{\epsilon/2}\int{d^dp\over (2\pi)^d}{1\over s+p^2}\right|_{s=\kappa^2}.
\end{equation}
The beta function is then given by 
$\beta(g_R)\equiv\kappa\partial_{\kappa}g_R=g_R^2/2\pi$ in $d=2$.
This is exact to all orders in $g_R$.
The field $\langle a\rangle$ now satisfies the renormalization group equation 
\begin{equation}
\left(\kappa{\partial\over\partial\kappa}+\beta(g_R){\partial\over\partial
g_R}\right)\langle a\rangle(x,g_R,\kappa)=0. \label{CS}
\end{equation}
The solution of this equation is given by
\begin{equation}
\langle a\rangle(x,g_R,\kappa)=(\kappa x)^{-d}\langle a
\rangle(\kappa^{-1},\tilde g_R,\kappa)\qquad (d\leq 2), 
\label{csz}
\end{equation}
where the running coupling $\tilde g_R$ acquires the universal asymptotic form
\begin{equation}
\tilde g_R\sim{2\pi\over\ln x}\left[1+O[(\ln x)^{-1}]\right]\qquad(d=2).
\label{rr2dasy}
\end{equation}

The leading order (tree) diagrams for the density are exactly equivalent to the
mean field rate equations \cite{lee}. 
Hence inserting the mean field expression (\ref{mfs}) into the RG 
equation (\ref{csz}) yields the exact leading asymptotic 
expression
\begin{equation}
\langle a(x)\rangle\sim {2\ln x\over\pi x^2}\left(1+O[(\ln x)^{-1}]\right)
\qquad (d=2). \label{mfRGs}
\end{equation}
We note that this renormalization procedure is only
needed at or below the upper critical dimension. For $d>2$ the mean field 
results in (\ref{mfs}) become qualitatively correct. 

The next step is to turn the expression (\ref{qcalc}) for the persistence 
probability into a cumulant expansion (see \cite{cardy2} for a similar case),
\begin{eqnarray}
& & s(q,L)\sim\left\langle\exp\left[-(1-q^{-1})\int
d^dx~a(x)\right]\right\rangle= \label{cep} \\
& & \hspace{-.3in} \exp\left[-(1-q^{-1})\int d^dx~\langle a(x)\rangle 
+{1\over 2}(1-q^{-1})^2\int~d^dx\int~d^dx'~\langle a(x)a(x')\rangle_c-
\cdots\right].\label{cumexp}
\nonumber
\end{eqnarray}
We now consider the behaviour of $s(q,L)$ in different dimensions.

\noindent $\bullet~d=2$.
We use (\ref{mfRGs}) to derive the leading 
behaviour of the first term in the cumulant expansion. 
Hence for a system of radius $L$, 
\begin{equation}
\int d^2x~\langle a(x)\rangle \sim \int d^2x~{2\ln x\over\pi x^2}
\sim 2(\ln L)^2, \label{mfRGnum}
\end{equation}
where the next order term is suppressed by $(\ln L)^{-1}$.
This result implies that almost all of the particles 
in the system are situated very far from the point of injection. This
confirms that 
the behaviour of the model close to the injection point is unimportant when
considering the large $L$ limit. 
Substituting (\ref{mfRGnum}) into (\ref{cep}) gives
\begin{eqnarray}
& & s(q,L)\sim\exp\left [-h_2(q)(\ln L)^2+h^{(2)}_2(q)\,\ln L+\cdots\right ]
\label{rql} \\
& & h_2(q)=2(1-q^{-1})+4C_2(1-q^{-1})^2+\cdots, \label{rqlconsts}
\end{eqnarray}
where we assume that all the higher order cumulants also 
have leading order contributions of ${\cal O}(\ln L)^2$.
Simulations of the reaction-diffusion system have confirmed this 
expectation (see below). Hence for the 
Voter model, using finite size scaling 
$L\sim t^{1/2}$ in (\ref{rql}), we have
\begin{eqnarray}
& & r(q,t)\sim\exp\left[-f_2(q)(\ln t)^2+f^{(2)}_2(q)\,\ln t+\cdots\right], 
\qquad (d=2) \label{rqt} \\
& & f_2(q)=(1/2)(1-q^{-1})+C_2(1-q^{-1})^2+\cdots. \label{rqtf}
\end{eqnarray}
This analysis shows that the unusual logarithmic form of this decay is due to
the fact that $d=2$ is a critical dimension.

\noindent $\bullet~2<d\leq 4$. Using the 
 mean field expressions 
(\ref{mfs}) yields the leading behaviour of the first term in 
the cumulant expansion
\begin{equation}
\int d^dx~\langle a(x)\rangle\propto\cases{
L^{d-2}/\lambda, & ($2<d<4)$\cr
L^2/(\lambda\ln L), & ($d=4$).} \label{mfnumb}
\end{equation}
Hence almost all of the 
particles are again situated far from the injection site.
Furthermore, dimensional analysis can be combined with an analysis of the 
diagrams for the higher order cumulants to show that each 
term in the cumulant expansion scales identically to (\ref{mfnumb}). 
We assume that a logarithm is present in each term of the cumulant 
expansion for $d=4$, although this analysis only explicitly shows the 
presence of a logarithm in the first term. 
Using finite size scaling, we 
then have
\begin{eqnarray}
& & r(q,t)\sim\cases{\exp\left[-f_d(q)t^{(d-2)/2}+\cdots\right], & ($2<d<4$)\cr
\exp\left[-f_4(q)t/\ln t+\cdots\right], & ($d=4$)} \\
& & f_d(q)=B_d(1-q^{-1})+C_d(1-q^{-1})^2+\cdots.
\end{eqnarray}
Note that the cumulant amplitudes
$B_d, C_d,\ldots$ are no longer universal.

\noindent $\bullet~d>4$. 
We again use (\ref{mfs}) for the first
term in the cumulant expansion,
\begin{equation}
\int d^dx~\langle a(x)\rangle\propto L^2. \label{dg4num}
\end{equation}
For $d>4$ the reaction process is asymptotically irrelevant since the
particles no longer ``see'' each other. Since the (non-Poissonian) 
reaction process is not then present, the diagrams for the higher order 
cumulants actually vanish. Hence we have
\begin{eqnarray}
& & r(q,t)\sim\exp[-f_d(q) t+\cdots] \qquad (d>4) \\
& & f_d(q)=B_d(1-q^{-1}).
\end{eqnarray}

In summary, we find that, at large times and in any dimension, 
$r(q,t)\sim\exp[-f(q)\langle m(t)\rangle]$, where $\langle m(t)\rangle$ is the
average  number of particles in the dual reaction--diffusion models at
time $t$ (see also \cite{DG}). 
Note that in $d=1$, $\langle  m(t)\rangle$ grows logarithmically in time, in
accord with the algebraic fall off of the persistence probability.

Let us finally report on simulations. 
We measured the persistence probability
$r(q,t)$ for the Voter model, as a function of time, with $q=2,\ldots,\infty$. 
The results for $d=2$ on a system of size $(1000)^2$
are shown in figure 1, where a decay $r(q,t)\sim\exp[-f_2(q)(\ln t)^2]$
is clearly seen, in agreement with both our predictions and the 
simulations for $q=2$ in \cite{bennaim}. A plot of $f_2(q)$ extracted 
numerically is given in figure 2. It shows qualitative agreement with 
the prediction given in (\ref{rqtf}).
In particular for increasing values of $q$, $f_2(q)$ progressively
departs from the first predicted term (dashed line on the figure).

Simulations performed on the $3d$ Voter model of size $(100)^3$
do not show the expected decay 
$r(q,t)\sim\exp[-f_3(q)t^{1/2}]$ at large times.
Instead, the numerical results indicate an effective exponent for $t$
slightly larger than 1/2. 
This is due to large subleading corrections which make
extraction of the leading behaviour difficult.

We now turn to simulations of the dual model I.
In $d=2$ we measured $\langle m\rangle=\int d^dx\langle a(x)\rangle$,
the average number of particles in the stationary state, for system sizes
$L=10,\ldots, 80$, averaged over $2\times 10^6$ MC steps (figure 3). 
The data are consistent with
$\langle m\rangle\sim 2(\ln L)^2$ plus corrections of ${\cal O}(\ln L)$,
in agreement with (\ref{mfRGnum}). 
This plot also provides evidence for the equivalence between 
models I and II, since our simulations are for model I whereas the
theory was developed for model II. 

We also examined the behaviour in $d=2$ of the first three higher order 
cumulants for the $a(x)$ field (integrated over the system size as in 
(\ref{cumexp})). 
They all scale with $L$ in the same way as the 
first cumulant $\int d^d x\langle a(x)\rangle=\langle m\rangle$.
This confirms that, in principle, each term in the cumulant expansion is 
equally ``relevant'' for the persistence calculation. 
Nevertheless the numerical prefactors for each term appear to decrease fairly
rapidly indicating that the first few terms are the most significant. 

Finally, in $d=3$ we measured $\langle m\rangle$ as a function of $L$,
for $L=10,\ldots,40$. 
For large sizes, instead of the prediction $\langle
m\rangle\sim L$ (see (\ref{mfnumb})), the data show an apparent
exponent of about $1.2$, indicating again the presence of large subleading
corrections.

\newpage
\begin{figure}
\begin{center}
\leavevmode
\vbox{
\epsfysize=3in
\epsffile{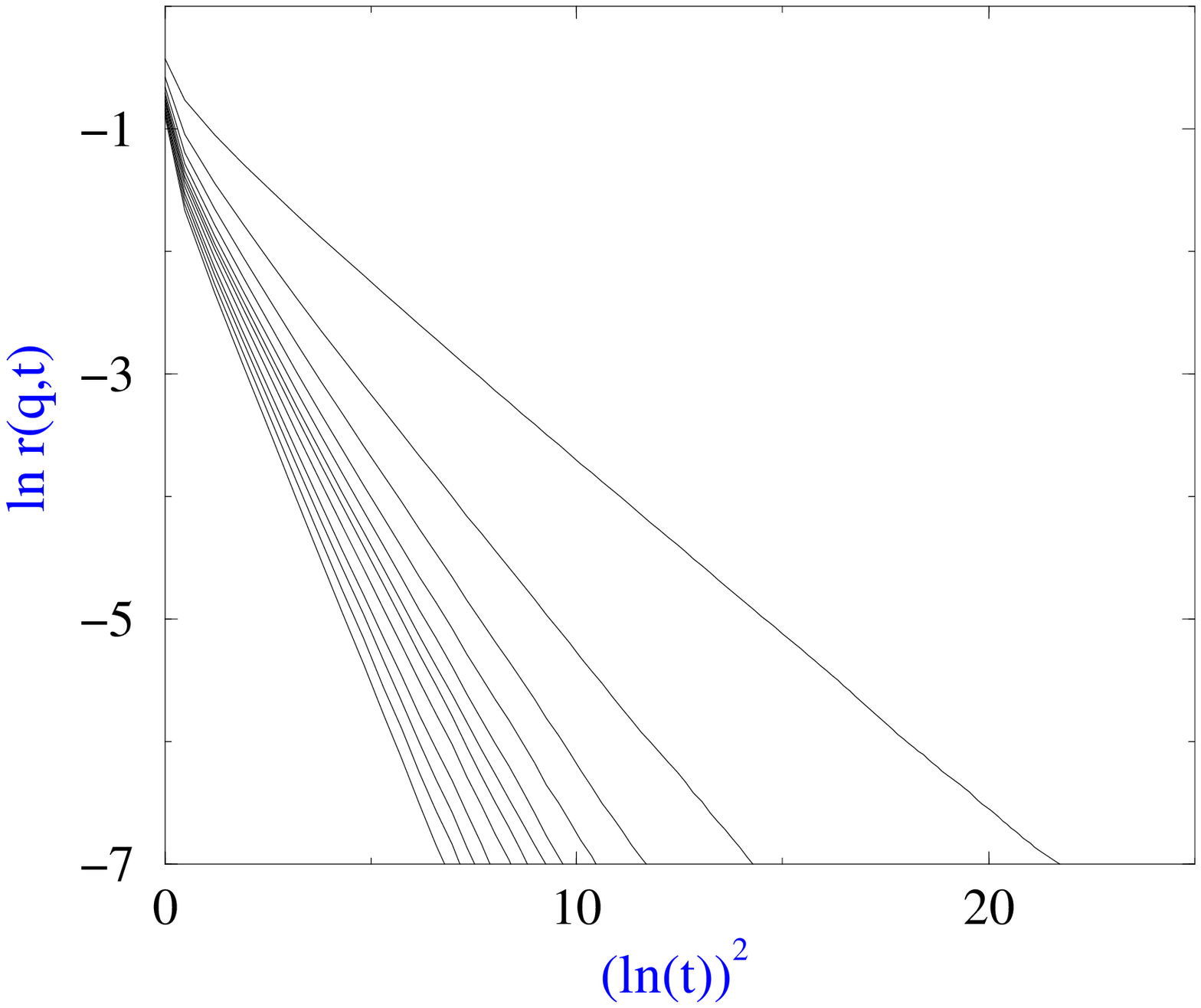}}
\end{center}
\caption{$\ln r(q,t)$ plotted against $(\ln t)^2$ for
$q=2$, 3, 4, 5, 6, 7, 8, 10, 14, 20, 40, $10^6$ 
(from top to bottom), in the
$2d$ Voter model (system size $(1000)^2$).} 
\end{figure}
\begin{figure}
\begin{center}
\leavevmode
\vbox{
\epsfysize=3in
\epsffile{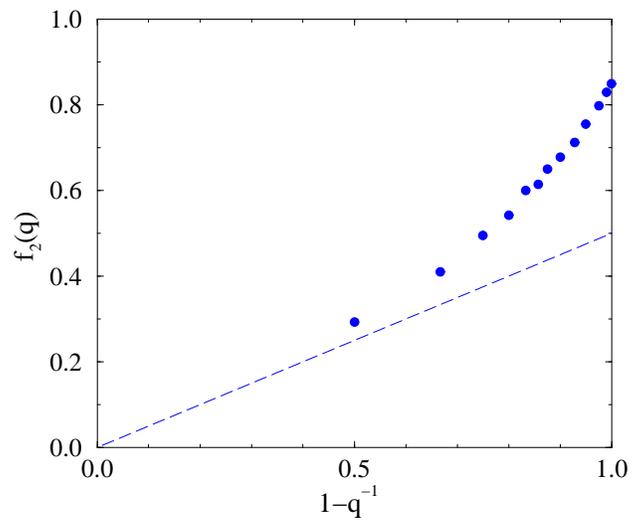}}
\end{center}
\caption{Plot of the numerical data for $f_2(q)$, extracted from figure 1,
 versus $(1-q^{-1})$ (dots). Also illustrated is
the lowest order theoretical approximation to $f_2(q)$ (dashed line).}
\end{figure}
\begin{figure}
\begin{center}
\leavevmode
\vbox{
\epsfysize=3in
\epsffile{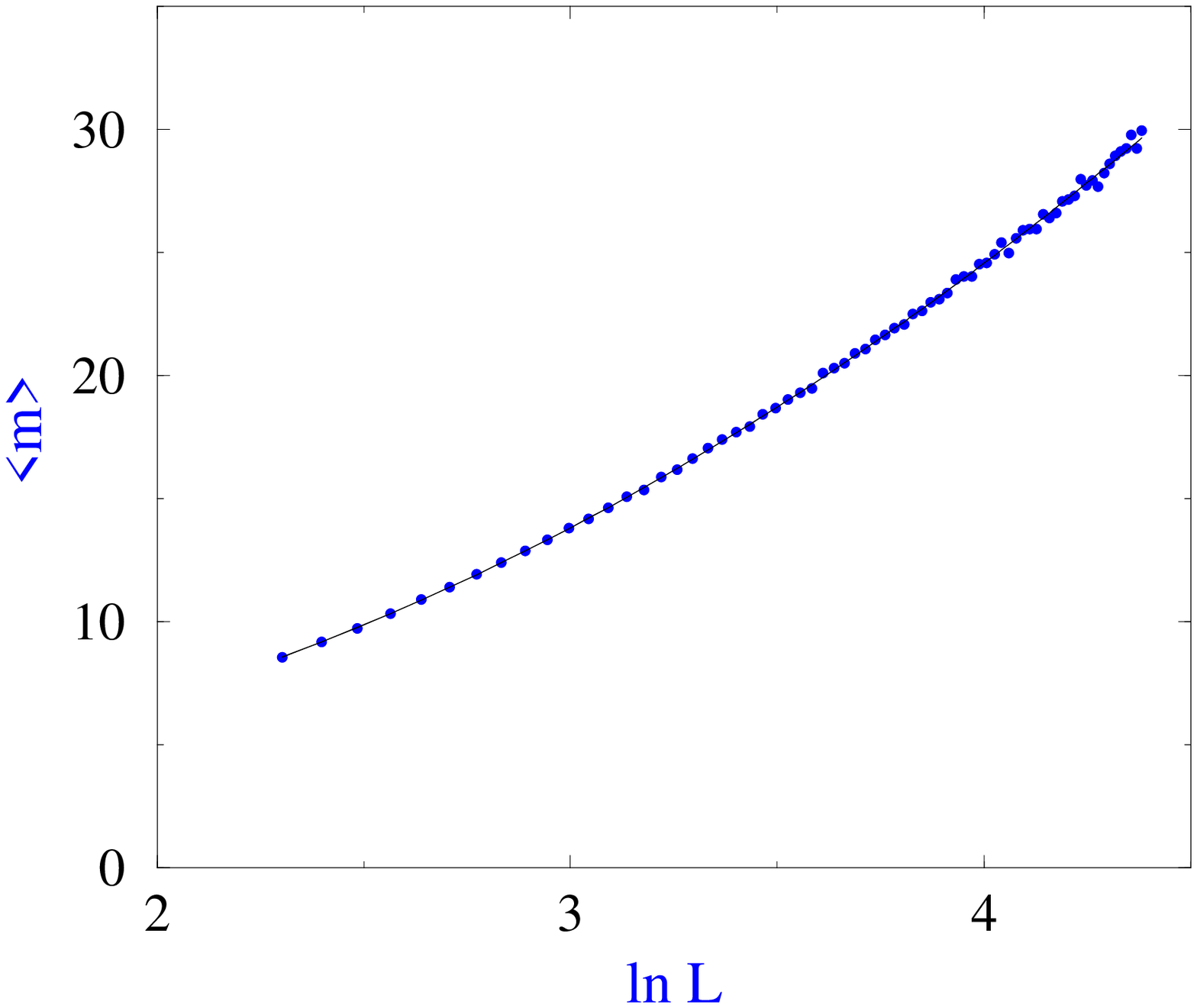}}
\end{center}
\caption{Plot of $\langle m\rangle$ versus $\ln L$ for system sizes $L=10,
\ldots,80$, averaged over $2\times 10^6$ MC steps (for model I). Solid line 
is a quadratic best fit to the data, where the coefficient of the quadratic 
term is found to be close to $1.9$.}
\end{figure}

\begin{thebibliography}{99}
\bibitem{Liggett} Liggett T M 1985 {\it Interacting Particle Systems} 
(Springer)
\bibitem{derrida1} Derrida B, Bray A and Godr\`eche C 1994 {\it J. Phys. A:
Math. Gen.} {\bf 27} L357
\bibitem{derrida2} Derrida B 1995 {\it J. Phys. A: Math. Gen.} {\bf 28} 1481
\bibitem{derrida3} Derrida B, Hakim V, Pasquier V 1995 {\it Phys. Rev. 
Lett.} {\bf 75} 751
\bibitem{bennaim} Ben-Naim E, Frachebourg L and Krapivsky P L 1996
{\it Phys. Rev. E} {\bf 53} 3078
\bibitem{cheng} Cheng Z, Redner S and Leyvraz F 1989 {\it Phys. Rev. Lett.}
{\bf 62} 2321
\bibitem{krapivsky} Krapivsky P L 1993 {\it Physica A} {\bf 198} 157
\bibitem{doi} Doi M 1976 {\it J. Phys. A: Math. Gen.} {\bf 9} 1465, 1479
\bibitem{peliti} Peliti L 1985 {\it J. Physique} {\bf 46} 1469
\bibitem{lee} Lee B P 1994 {\it J. Phys. A: Math. Gen.} {\bf 27} 2633
\bibitem{howard1} Howard M and Cardy J 1995 {\it J. Phys. A: Math. Gen.}
{\bf 28} 3599
\bibitem{cardy2} Cardy J 1995 {\it J. Phys. A: Math. Gen.} {\bf 28} L19
\bibitem{DG} Dornic I and Godr\`eche C, in preparation
\end{thebibliography}
\end{document}